\documentclass{article}

\usepackage{algorithm}
\usepackage[noend]{algpseudocode}
\usepackage{amsmath}
\usepackage{amsfonts}
\usepackage{amssymb}
\usepackage{mathtools}
\usepackage{graphicx}
\usepackage{textcomp}
 \usepackage[table]{xcolor}
\begin{document}





\huge{Digital synthesis of multistage etalons for enhancing the FSR}\\

\normalsize{Faiza Iftikhar$^1$ Usman Khan$^2$ and M. Imran Cheema$^{1,*}$}\\

{$^1$Department of Electrical Engineering, Syed Babar Ali School of Science and Engineering, LUMS, Sector U, DHA, Lahore, Pakistan\\
$^2$Electrical and Computer Engineering, Tufts University Medford, MA 02155, USA\\

$^*$imran.cheema@lums.edu.pk\\ 


\begin{abstract}
Fabry-Perot fiber etalons (FPE) built from three or more reflectors are attractive for a variety of applications including communications and sensing. For accelerating a research and development work, one often desires to use off-the-shelf components to build an FPE with a required transmission profile for a particular application. Usually, multistage FPEs are designed with equal lengths of cavities followed by determination of the required reflectivities for realizing a desired transmission profile. As seen in previous works, fabricated reflectors are usually slightly different from the designed ones leading to departure from the desired transmission profile of the FPE. Here, we show a novel digital synthesis of multistage etalons with off-the-shelf reflectors and unequal lengths of involved cavities. We find that, in contrast to equal cavity lengths, unequal lengths of cavities provide more number of poles in the $z$-domain to achieve a desired multicavity FPE transmission response. For given reflectivities and by determining correct unequal lengths of cavities with our synthesis technique, we demonstrate a design example of increasing the FSR followed by its experimental validation. This work is generalizable to ring resonators, mirrored, and fiber Bragg grating based cavities; enabling the design and optimization of cavity systems for a wide range of applications including lasers, sensors, and filters.
\end{abstract}

\section{Introduction}
The transmission profile of multistage Fabry-Perot etalons (FPEs) can be manipulated as a function of the involved reflectivities and lengths of cavities. This control of transfer function has enabled researchers to utilize these multicavity structures in various communications applications.  Traditionally, researchers determine reflectivities for an FPE system by assuming equal lengths of cavities for a desired application \cite{Zhang:17,pruessner2013,Cheng:13,bae2005,yim2005,yim2004}. By looking at the reverse problem i.e., determination of lengths of cavities for given reflectivities to realize a particular application has not been fully explored yet.

In previous works, the concept of unequal cavity FPEs and its implications in terms of transmission response was first introduced by Stadt et al. using a matrix approach\cite{vandeStadt:85}. In order to deal with complexity of the non-linear transfer function of multistage FPEs, researchers came up with a $z$-domain technique for the design and analysis of optical structures \cite{MacFarlane:94}. In the aforementioned works, the lengths of cavities were selected randomly to demonstrate their techniques. The $z$-domain techniques have been utilized for designing a variety of multistage FPE systems including interferometers \cite{Zhang:10, Cheng:13}, interleavers \cite{Cao:04,Zhang:17}, and filters \cite{Madsen_1998,1994JLwT...12..471D,ahmed2011}

The multistage FPEs with equal cavities require determination of reflectivities for realizing a required transmission response. The implemented FPE system has always different transmission response due to fabrication tolerances of the reflectors \cite{Zhang:17}. During the implementation stage, although one can not do much about the fabricated reflectors however, one can still control the lengths of cavities especially in mirror and FBG based FPEs. Considering all these issues, we propose a novel digital synthesis technique for determining optimum lengths of cavities for a given set of reflectivities to achieve a desired transmission response of multistage FPEs.

Our design process starts with a given transmission profile. We then estimate a $z$-domain transfer function from the given transmission response by using a vector fitting technique \cite{Ozdemir_2017}. With our developed algorithm, we then map the estimated transfer function to an FPE system with a desired number of reflectors and given reflectivities. This process provides lengths of cavities to achieve the desired transmission response with the given set of reflectivities. We show the application of our proposed synthesis technique by providing modeling and experimental results of FPEs comprising of two, three, and four cavities to achieve a desired free spectral range (FSR) and peak rejection ratio (PR) of side bands in a given transmission response.

We now describe the rest of the paper. In Section \ref{sec:z_gen}, we provide an analytical expression for finding $z$-transform of an $n$ cavity FPE. We provide our digital synthesis technique in Section \ref{sec:DSP} followed by theoretical and experimental validation of the technique with a design example in Section \ref{sec:des_eg}. Finally, we provide concluding remarks in Section \ref{sec:con}.

\section{Generalized $z$-domain transmission function of a multistage FPE}\label{sec:z_gen}
The schematics of a multistage FPE is shown in the Fig.\ref{fig:MFPE}. Each cavity is assumed to be formed by connecting fiber Bragg gratings (FBGs) through optical fibres of unequal lengths. After invoking the Stokes parameters, $r_n^- = -r_n^+$ , $t_n^+t_n^-+r_n^+r_n^- = 1$ and the matrix approach outlined in \cite{vandeStadt:85}, it can be shown for the multistage FPE system in Fig.\ref {fig:MFPE},
\begin{figure}[htbp]
\centering
\fbox{\includegraphics[width=0.9\linewidth]{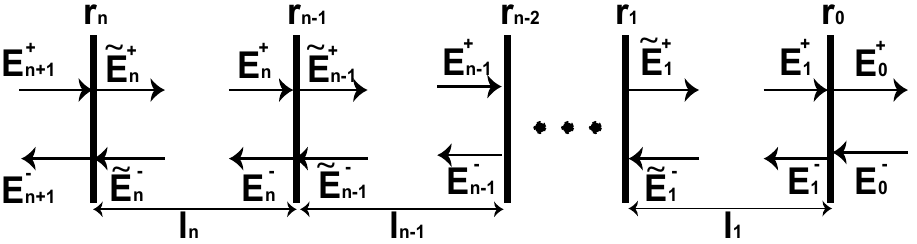}}
\caption{Multistage FPE structure with $n$ cavities. $E$ represents an electric field, $\Tilde{E}$ represents a phase shifted electric field due to a cavity length $l$, and  $r$ represents the peak amplitude reflection coefficient of an FBG.}
\label{fig:MFPE}
\end{figure}

\begin{equation}\label{eq:16}
\begin{aligned}
\begin{bmatrix}E_{n+1}^+ \\E_{n+1}^- \end{bmatrix}=\frac{e^{j\sum\limits_{m=1}^n\phi_m}}{t_n...t_3.t_2.t_1.t_0}\begin{bmatrix}1 & r_ne^{-j2\phi_n} \\r_n & e^{-j2\phi_n} \end{bmatrix}........\\
\begin{bmatrix}1 & r_2e^{-j2\phi_2} \\r_2 & e^{-j2\phi_2} \end{bmatrix}.\begin{bmatrix}1 & r_1e^{-j2\phi_1} \\r_1 & e^{-j2\phi_1} \end{bmatrix}.\begin{bmatrix}1 & r_0 \\r_0 & 1\end{bmatrix}\begin{bmatrix}E_0^+ \\E_0^- \end{bmatrix}
\end{aligned}
\end{equation}

\begin{equation}\label{eq:17}
\begin{aligned}
\begin{bmatrix}E_{n+1}^+ \\E_{n+1}^- \end{bmatrix}=\frac{e^{j\sum\limits_{m=1}^n\phi_m}}{\prod\limits_{i=0}^nt_i}\begin{bmatrix}A_n & B_n  \\C_n & D_n \end{bmatrix}\begin{bmatrix}E_0^+ \\E_0^- \end{bmatrix}
\end{aligned}
\end{equation}

where $\phi_n=\dfrac{2 \pi n_n l_n}{\lambda_o}$ represents phase length of the $n^{th}$ cavity, $\lambda_o$ is the source wavelength, $n_nl_n$ is the optical path length of the $n^{th}$ cavity, $r_n$ and $t_n$ are amplitude reflection and transmission coefficients at the $n^{th}$ FBG, respectively and `$+$`, `$-$' indicate directions of fields towards the right and left, respectively. By applying the boundary conditions, we then determine the transmission amplitude of the $n$ cavity FPE system,

\begin{equation}\label{eq:18}
t_n = \frac{E_0^+}{E_{n+1}^+}\mid_{E_0^- = 0}=\frac{e^{-j\sum\limits_{m=1}^n\phi_m}\prod\limits_{i=0}^nt_i}{a_n}
\end{equation}

After analyzing above equations, we see a definite pattern of terms appearing in $a_n$ and its deduced expression is given by
\begin{equation}\label{eq:18a}
a_n = 1+\sum_{k=1}^{2^n-1}C_{{P}(n)[k]} e^{-j2\sum\limits_{m \in {P}(n)[k]}^{\#P(n)[k]} \phi_m}
\end{equation}
where $P(n)$ is the power set of $1$ to $n$ cavities excluding the empty set, $P(n)[k]$ is $k^{th}$ subset of the $P(n)$, $\#P(n)[k]$ is the cardinality of the subset $P(n)[k]$, and $C$ is a constant associated with each subset. The constant $C$ is made-up of reflectivities of an individual cavity or combination of cavities. As an example for a three cavity FPE (i.e., 4 FBGs), $P(n=3)=\left\{\{1\},\{2\},\{3\},\{1,2\},\{1,3\},\{2,3\},\{1,2,3\}\right\}$, $P(n)[k=4]=\{1,2\}$, $\#P(n)[k=4]=2$, $C_{{P}(3)[1]}=r_or_1$, $C_{{P}(3)[2]}=r_1r_2$, $C_{{P}(3)[3]}=r_2r_3$, $C_{{P}(3)[4]}=r_or_2$, $C_{{P}(3)[5]}=r_or_1r_2r_3$, $C_{{P}(3)[6]}=r_1r_3$, $C_{{P}(3)[7]}=r_or_3$. The expressions for $C$ follow a specific set of rules: a)The $C$ of one cavity will have product of reflection coefficients spanned by that cavity b)The $C$ of two joint cavities will have product of reflection coefficients at their ends c)The $C$ of two disjoint cavities will have product of all reflection coefficients spanned by each cavity. Physically, these rules make perfect sense as all individual cavities and their combinations (coupling) will dictate the overall transmission response of the multistage FPE.

On the same lines as outlined in \cite{MacFarlane:94}, we can transform the transmission response given by Eq. \eqref{eq:18}  and our derived Eq. \eqref{eq:18a} in the $z$-domain as

\begin{equation}\label{eq:19}
t_n(z) =\frac{z^{-\sum\limits_{m=1}^nx_m}\prod\limits_{i=0}^nt_i}{A_n}
\end{equation}

\begin{equation}\label{eq:19a}
A_n = 1+\sum_{k=1}^{2^n-1}C_{{P}(n)[k]}. z^{\scriptstyle -2\sum\limits_{m \in {P}(n)[k]}^{\#P(n)[k]} x_m}
\end{equation}
where $z^{-2x_m }= e^{-j\dfrac{4\pi \text{n}_m \text{l}_m}{\lambda_o}}$. By using equations \eqref{eq:19} and \eqref{eq:19a}, we can directly get the $z$-domain transmission function of an FPE system for any number of reflectors and lengths without invoking any algebraic manipulations.

It is evident from Eq. \eqref{eq:19a} that $n$ equal cavities are equivalent to adding $n$ number of poles to the system whereas $n$ unequal cavities are equivalent to $2^n-1$ poles. This makes the unequal cavity system more powerful in achieving the desired transmission response.
\section{Digital synthesis procedure}\label{sec:DSP}
The overall procedure for digitally synthesizing a required FPE is shown in Fig. \ref{fig:DSP}.
\begin{figure}[htbp]
\centering
\fbox{\includegraphics[width=0.9\linewidth]{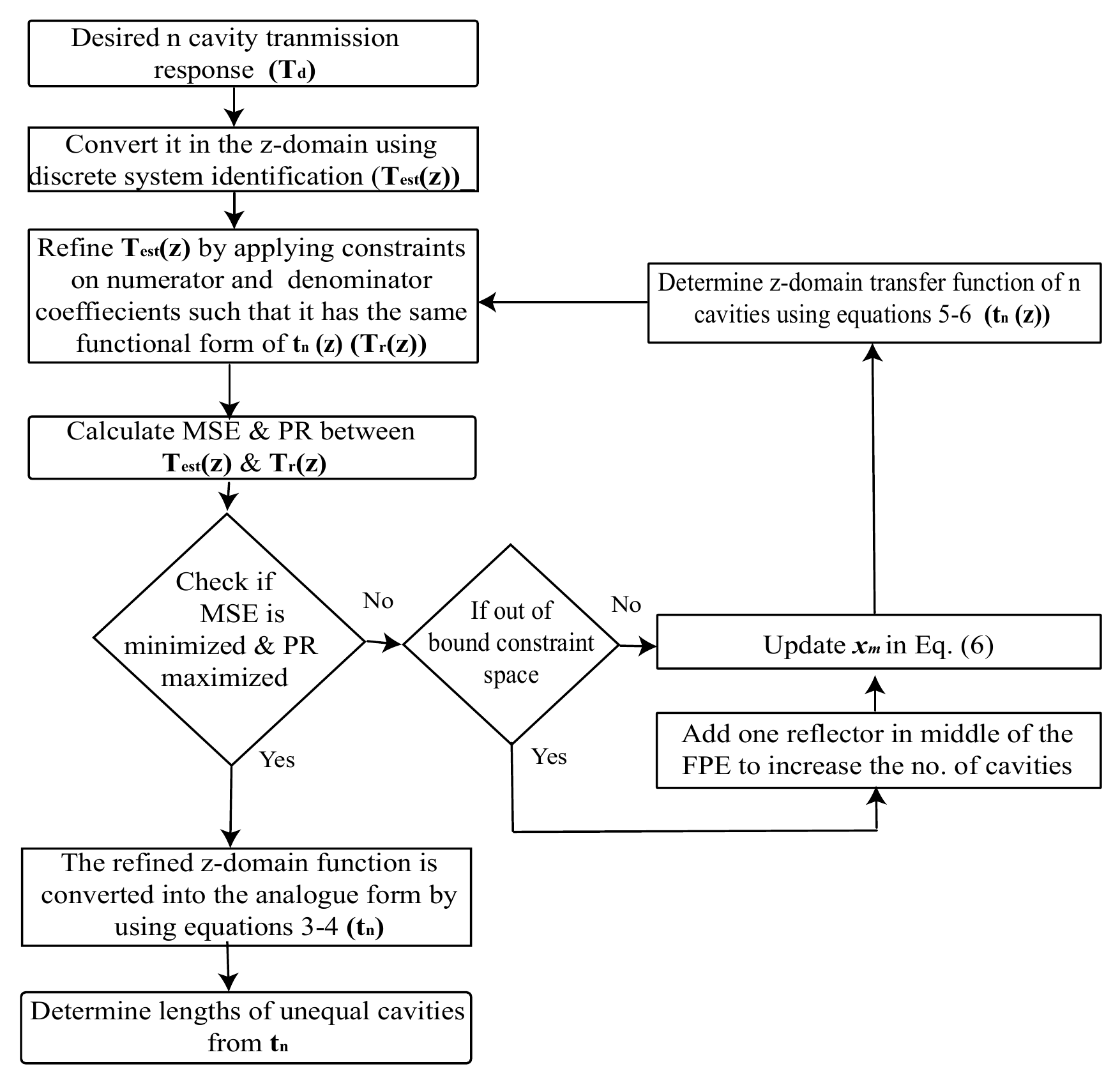}}
\caption{Block diagram for digital synthesis of an FPE with the desired transmission profile. MSE-Mean square error, PR-Peak stopband rejection}
\label{fig:DSP}
\end{figure}
Starting from a desired transmission response (T$_d$), we use discrete system identification via vector fitting \cite{Ozdemir_2017} to estimate its transfer function T$_{est}(z)$ in the $z$-domain. The obtained T$_{est}(z)$ is generally a higher order function with typically more than 100 poles and zeros. With the aim to map T$_{est}(z)$ to $n$ cavity system with unequal lengths of cavities, we refine the T$_{est}(z)$ by applying constraints on coefficients of the denominator and numerator polynomials. For this purpose, we determine the $z$-domain transmission function, t$_n(z)$, of the $n$ cavity etalon by using equations \eqref{eq:19}-\eqref{eq:19a} for a given set of reflectivities.  We calculate coefficient constraints based on fixed reflectivities of reflectors. We elaborate on the constraint procedure in the design example and appendix. The estimated transfer function T$_{est}(z)$ is refined iteratively using the predictive error method (PEM) \cite{Pintelon_2012} to achieve the same functional form of t$_n(z)$. The refined function is called as T$_r(z)$. During the refinement process at each iteration, we determine the mean square error (MSE) and peak stopband rejection (PR) by comparing T$_{est}(z)$  and T$_r(z)$. If we do not get a required MSE and PR in a given constraint set of $n$ cavity system, one more reflector is added in the middle of the system to increase the number of cavities. When we get the required T$_r(z)$  with the minimum MSE and maximum PR, we convert it into the analogue form, t$_n$,  using equations \eqref{eq:18}-\eqref{eq:18a}. We then determine the respective lengths of $n$ cavities.

\section{Design example}\label{sec:des_eg}
We now present modelling and experimental results of enhancing the FSR and PR of 2, 3, and 4 cavity FPE systems by determining the lengths of cavities using the digital synthesis procedure described in Section \ref{sec:DSP}.
\subsection{Modelling results}\label{sec:ModR}
Consider that we are given a two cavity system at 1550nm with fixed reflectivities of $R_0 = 0.87, R_1 = 0.99, R_2 = 0.91$ and equal optical path lengths of $n_1l_1$ = $n_2l_2$=90cm. Although these specifications can be picked somewhat arbitrarily however, we pick these numbers by keeping our experiments in mind (see Section \ref{sec:ExpR}). We choose these reflectivities due to availability of respective fiber Bragg gratings (FBGs) in our lab. The optical path lengths are selected due to easiness of splicing of two FBGs. By using equations \eqref{eq:18}-\eqref{eq:18a}, we can obtain the transmission of the given structure as shown in Fig. \ref{fig:two cavity structure}a.

The free spectral range (FSR) of the given structure is 1.344pm. Now consider that we require to increase the FSR by 15 times i.e., to 20pm with the peak stopband rejection of better than -30dB. We process the two equal cavity transmission response to suppress peaks in such a way that the FSR is enhanced by 15 times as shown in Fig. \ref{fig:two cavity structure}b. During the processing, we take 15 FSRs of the given transmission profile and apply a digital bandpass filter with PR of -40dB to the central peak in Matlab.

\begin{figure}[htbp]
	\centering
	\fbox{\includegraphics[width=0.9\linewidth]{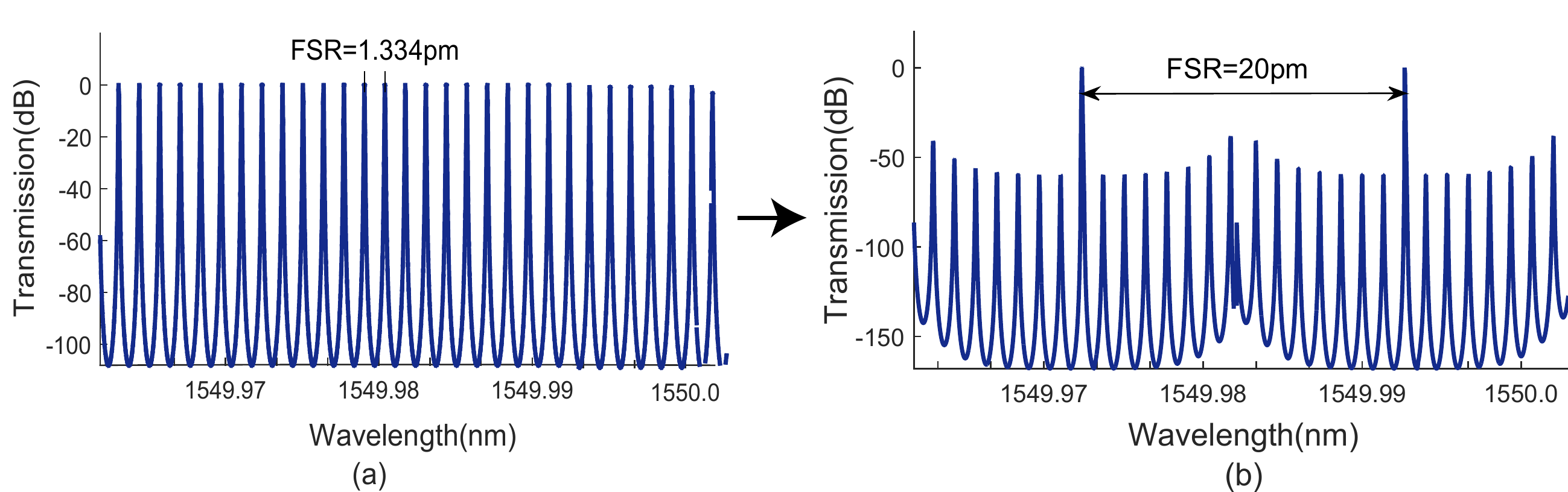}}
	\caption{Processing of Transmission profile of two-equal-cavity system to enhance FSR.}
	\label{fig:two cavity structure}
\end{figure}

After generating the desired transmission profile, t$_d$, we use \texttt{tfest} method in Matlab \cite{Ozdemir_2017} to estimate its $z$-domain transfer function, T$_{est}(z)$. While estimating the transfer function we keep on increasing the number of poles until the maximum percentage fit and minimum mean square error(MSE) are achieved as shown in Fig. \ref{fig:MSE}. For the present example, the estimated transfer function, T$_{est}(z)$, with 99.96\% match is of order 600.

\begin{figure}[htbp]
	\centering
	\fbox{\includegraphics[width=0.6\linewidth]{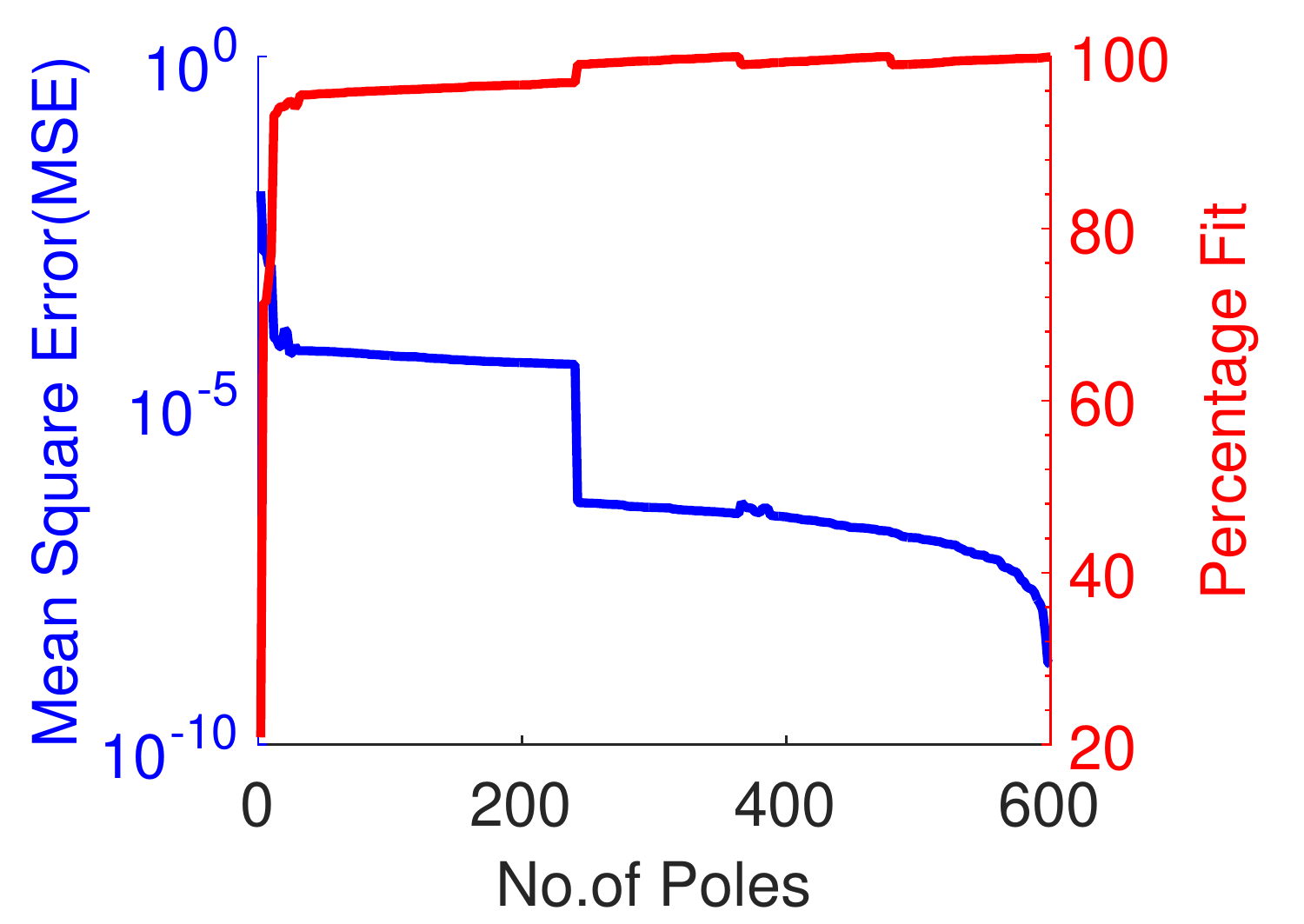}}
	\caption{MSE and Percentage fit analysis for estimating $z$-domain transfer function of the desired transmission response.}
	\label{fig:MSE}
\end{figure}

Now let us try to map T$_{est}(z)$ on a 2 cavity FPE system with unequal lengths whose $z$-domain transfer function, t$_n(z)$, is given by equations \eqref{eq:19}-\eqref{eq:19a}.  As more than 2$^2$=4 terms are non-zero in denominator polynomial of the transfer function, T$_{est}(z)$, we refine it by imposing fixed constraints on the required four coefficients, $C_{{P}(2)[1]}$,$C_{{P}(2)[2]}$,$C_{{P}(2)[3]}$, in the denominator and one coefficient, $t_0t_1t_2$ in the numerator. These constraints depend on values of $x_1$ and $x_2$ as elaborated in the appendix. This process gives us a refined transfer function, T$_r(z)$.

During the refinement process, we use the predictive error method (PEM) \cite{Pintelon_2012} which records MSE between T$_{est}(z)$ and T$_r(z)$ for each set of constrained coefficients. During the refinement procedure, we impose constraints on the aforementioned selected coefficients while forcing the rest of coefficients towards zero and keep updating the cavity lengths, $x_m$. We scan cavity lengths from ($x_1,x_2$)=(1cm,1cm) to ($x_1,x_2$)=(90cm,90cm) during the refinement. After the refinement step completion, we transform T$_r(z)$ to t$_n$ by using equations \eqref{eq:18}-\eqref{eq:18a} and determine lengths of cavities. Some of results obtained from the refinement process are shown in Table \ref{tab:2C}. For calculating lengths $l_1$ and $l_2$ we assume refractive index of 1.445 due to usage of SMF-28 fibers for our experimental results. Although the second combination in Table \ref{tab:2C} provides maximum PR however, we pick the shaded combination due to easiness of splicing.
\begin{table}[!htbp]\caption {Few representative refinement results including an optimum one for the two cavity FPE. All lengths are in centimeters.}\label{tab:2C}
	\begin{center}
	\begin{tabular}{@{}|l|l|l|l|l|l|@{}}
	\hline
{$x_1$} & {$x_2$} & {MSE} & {$l_1$} & {$l_2$} & {PR(dB)} \\ \hline
		 90 & 6 & 7.9 $\times$ 10$^{-7}$ & 62.07 & 4.14 & -16.52 \\\hline
		90 & 42 & 6.54 $\times$ 10$^{-7}$ & 62.07 & 28.96 & -19.29 \\\hline		
		\rowcolor{lightgray} 90 & 66 & 6.01 $\times$ 10$^{-7}$ & 62.07 & 45.52 & -17.75 \\ \hline
         90 & 78 & 6.81 $\times$ 10$^{-7}$ & 62.07 & 53.79 & -14.8 \\ \hline
	\end{tabular}
\end{center}
\end{table}

From the two cavity results, it is clear that the desired PR is not achievable. Therefore, we increase the number of cavities to achieve the required PR. Tables \ref{tab:3C} and \ref{tab:4C} show improvements in the PR corresponding  to three and four cavity FPEs, respectively. Again we pick the shaded combinations to avoid splicing challenges while connecting FBGs during experiments.  The modelling results for transmission of two, three, and four cavity FPEs are shown in Fig. \ref{fig:mod_res} for the obtained unequal lengths from our synthesis technique. For the three cavity FPE, we assume $R_0 = 0.87$, $R_1 = R_2 = 0.99$, and $R_3 = 0.91$. For the four cavity FPE, we assume $R_0 = 0.87$,  $R_1 = R_2 = R_3= 0.99$ and $R_4 = 0.91$. We have picked these reflectivities as per availability in our lab and for easier progression from two to three and then to the four cavity FPE system during experiments. 
\begin{figure}[!htbp]
	\centering
	\fbox{\includegraphics[width=0.7\linewidth]{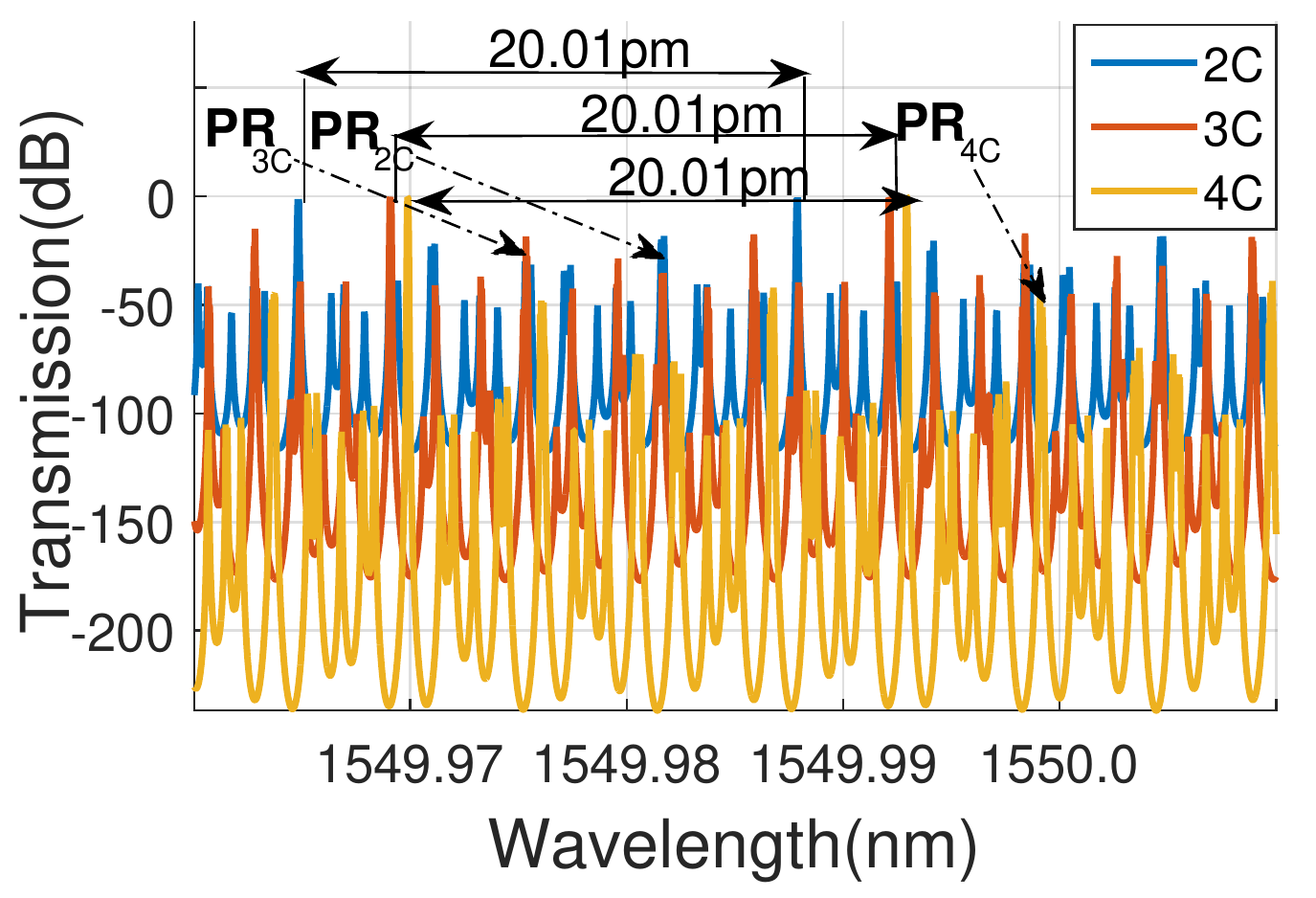}}
	\caption{Modeling results for transmission profiles of two, three, and four cavity FPEs plotted by using equations \eqref{eq:18}-\eqref{eq:18a}. The lengths used in these plots are shown by shaded rows of Tables \ref{tab:2C}, \ref{tab:3C}, and \ref{tab:4C}. C-Cavity.}
	\label{fig:mod_res}
\end{figure}

 \begin{table}[!htbp]\caption {Few representative refinement results including an optimum one for the three cavity FPE. All lengths are in centimeters.}\label{tab:3C}
 	\begin{center}
 		\begin{tabular}{@{}|l|l|l|l|l|l|l|l|@{}}
 		\hline
 			{$x_1$} & {$x_2$} & {$x_3$} & {MSE} & {$l_1$} & {$l_2$} &  {$l_3$} & {PR(dB)} \\ \hline
  	
 	 		90 & 6 & 6 & 2.02 $\times$ 10$^{-7}$ & 62.07 & 4.14 & 4.14 & -45.34 \\\hline
 	 		90 & 42 & 42 & 2.1 $\times$ 10$^{-7}$ & 62.07 & 28.96 & 28.96 & -21.94 \\\hline		
 	 		\rowcolor{lightgray} 90 & 66 & 66 & 2.22 $\times$ 10$^{-7}$ & 62.07 & 45.52 & 45.52 & -15.79 \\ \hline
 	 		90 & 78 & 78 & 2.28 $\times$ 10$^{-7}$ & 62.07 & 53.79 & 53.79& -12.99 \\ \hline

 	 	\end{tabular}
  	\end{center}
 	 \end{table}

\begin{table}[!htbp]\caption {Few representative refinement results including an optimum one for the four cavity FPE. All lengths are in centimeters.}\label{tab:4C}
	\begin{center}
		\begin{tabular}{@{}|l|l|l|l|l|l|l|l|l|l|@{}}
			\hline
			{$x_1$} & {$x_2$} & {$x_3$}  & {$x_4$}& {MSE} & {$l_1$} & {$l_2$} &  {$l_3$} & {$l_4$} & {PR(dB)} \\ \hline
			
			90 & 6 & 6 & 6& 9.3 $\times$ 10$^{-7}$ & 62.07 & 4.14 & 4.14 & 4.14 & -70 \\\hline
			90 & 90 & 42 & 42 & 9.98 $\times$ 10$^{-7}$ & 62.07 & 62.07& 28.96 & 28.96 & -40 \\\hline		
			\rowcolor{lightgray} 90 & 90 & 66 & 66 & 1.03 $\times$ 10$^{-7}$ & 62.07  & 62.07 & 45.52 & 45.52 & -37.49 \\ \hline
			90 & 90 & 102& 102& 1.15 $\times$ 10$^{-7}$ & 62.07 & 62.07 & 70.34 & 70.34 & -34 \\\hline
			
		\end{tabular}
	\end{center}
\end{table}

 \subsection{Experimental results}\label{sec:ExpR}
 The experimental schematics are shown in Fig. \ref{fig:exp_set}.
 \begin{figure}[!htbp]
	\centering
	\fbox{\includegraphics[width=0.9\linewidth]{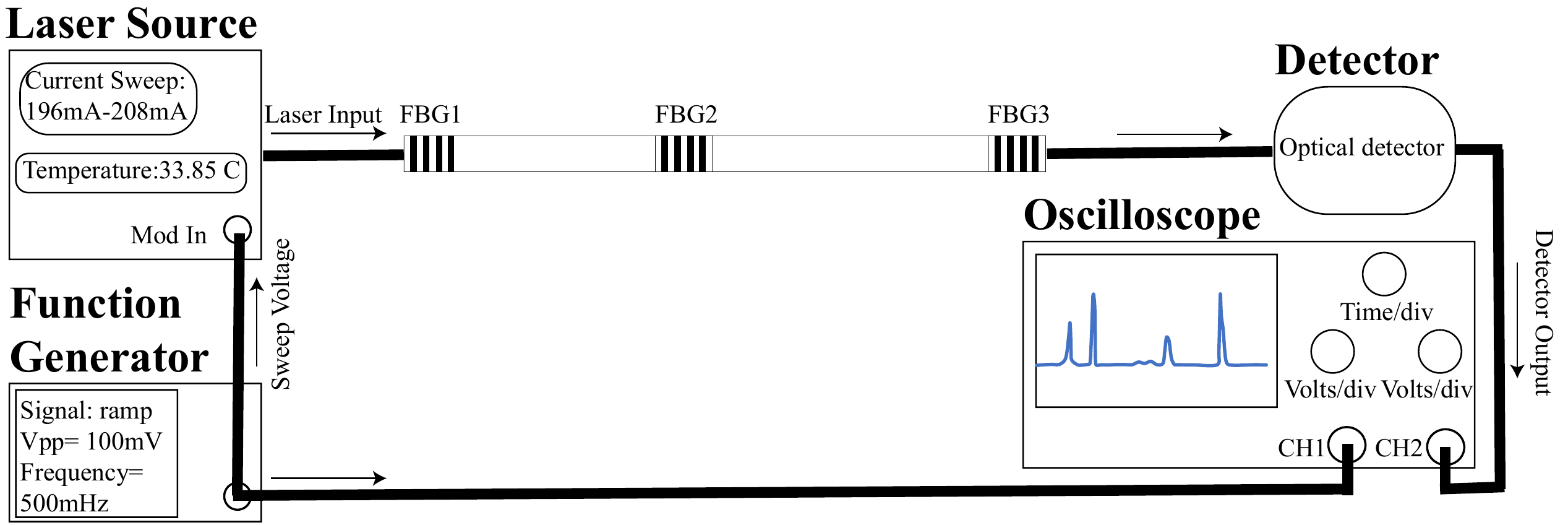}}
	\caption{Experimental Schematics.}
	\label{fig:exp_set}
\end{figure}
 In our experimental setup, we used an Eblana Photonics Inc. 1550nm laser diode (EP1550-0-NLW-B26-100FM) mounted on a Thorlabs driver (CLD1015). We tuned the laser diode using current modulation by employing a 0.5Hz triangular wave of 100mV$_{pp}$. We applied the laser signal to two, three, and four multicavity structures whose parameters are given in Section \ref{sec:ModR}. The output is displayed on an oscilloscope via Thorlabs detector (DET08CFC) followed by the collected data analysis in Matlab.  
 The experimental results are shown in Fig. \ref{fig:exp_res} which are in excellent agreement with the modeling results provided in Fig. \ref{fig:mod_res}.
\begin{figure}[!htbp]
 	\centering
 	\fbox{\includegraphics[width=0.7\linewidth]{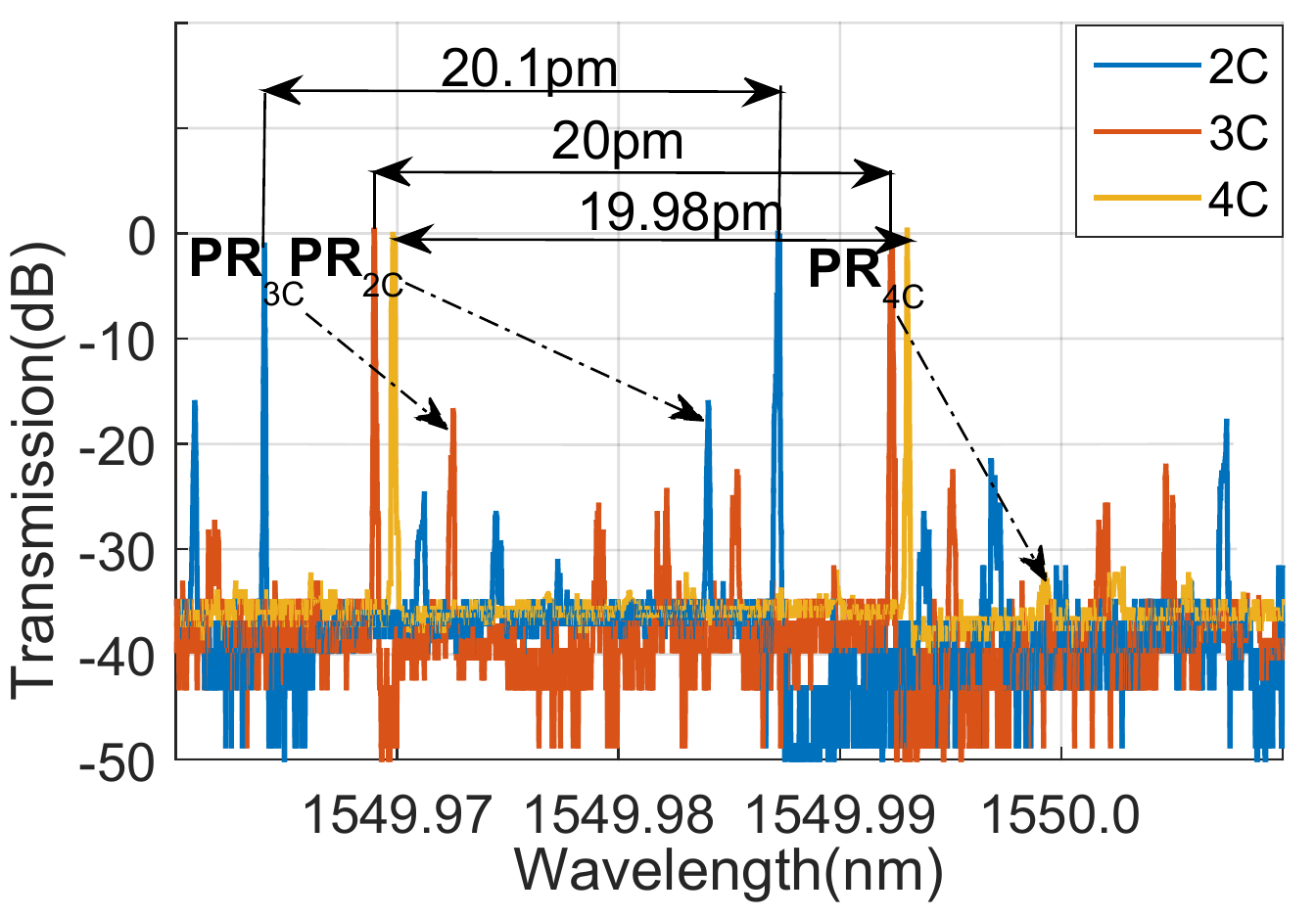}}
 	\caption{Experimental results for normalized transmission profiles of two, three, and four cavity FPEs. The lengths used in these plots are shown by shaded rows of Tables \ref{tab:2C}, \ref{tab:3C}, and \ref{tab:4C}. FBGs' reflectivities for each FPE are provided in Section \ref{sec:ModR}. C-Cavity.}
 	\label{fig:exp_res}
 \end{figure}
\section{Conclusion}\label{sec:con}
The modeling and experimental results clearly show that our proposed digital synthesis technique is achieving the desired transmission response using multistage FPEs. We find that the use of unequal cavities provide more number of poles in achieving a desired FPE transmission response as compared to the equal cavity approach. In the present work, we have demonstrated a controlled enhancement of the FSR by finding optimum lengths of cavities in multistage FPEs. We find that the FSR can be increased by any factor with two cavity etalon however, arbitrary increase in the FSR occurs at the expense of reduction in PR. For achieving both high FSR and PR, we need to add more FBGs.

Our digital synthesis approach can be highly useful in applications where one can employ off-the-shelf reflectors and vary inter-cavity lengths to achieve a desired transmission response. In the present work, although we present centimeter range cavities for demonstrating experimental results however, our digital synthesis technique is applicable to any length scale. We anticipate that the present work will find wide applications in developing filters, interleavers, sensors, and single mode lasers.

\section*{Funding}
Higher Education Commission of Pakistan (NRPU-5927).
\section*{Appendix}\label{app:app1}
Let us consider a case of the two cavity system with $x_1=90$cm and $x_2=1$cm. Using  equations \eqref{eq:19} and \eqref{eq:19a} we get
\begin{equation}\label{eq:app2}
t_2(z) = \frac{t_0t_1t_2z^{-91}}{1+C_{{P}(2)[1]}z^{-180}+C_{{P}(2)[2]}z^{-2}+C_{{P}(2)[3]}z^{-182}}
\end{equation}
where $C_{{P}(2)[1]}=r_0r_1$, $C_{{P}(2)[2]}=r_1r_2$, $C_{{P}(2)[3]}=r_0r_2$. In the design example, T$_{est}(z)$ has 600 poles therefore, denominator constraints for the 600 coefficients, $a_1-a_{600}$, are $a_1=1$, $a_2= C_{{P}(2)[2]}$, $a_{180}= C_{{P}(2)[1]}$, $a_{182}= C_{{P}(2)[3]}$, and the rest of coefficients in the denominator are forced to zero. Similarly, constraints on coefficients of the numerator, $b_1-b_{300}$, are such that only $b_{91}=t_0t_1t_2$ is non-zero while remaining coefficients are equal to zero. This procedure is repeated by iteratively changing values of $x_1$ and $x_2$ until the desired MSE and PR are achieved.

\bibliographystyle{plain}
\bibliography{references}

\begin{thebibliography}{10}

\bibitem{ahmed2011}
Osman~S Ahmed, Mohamed~A Swillam, Mohamed~H Bakr, and Xun Li.
\newblock Efficient design optimization of ring resonator-based optical
  filters.
\newblock {\em Journal of Lightwave Technology}, 29(18):2812--2817, 2011.

\bibitem{bae2005}
Jinho Bae, Joohwan Chun, and Thomas Kailath.
\newblock The schur algorithm applied to the design of optical multi-mirror
  structures.
\newblock {\em Numerical linear algebra with applications}, 12(2-3):283--292,
  2005.

\bibitem{Cao:04}
S.~Cao, J.~Chen, J.~N. Damask, C.~R. Doerr, L.~Guiziou, G.~Harvey, Y.~Hibino,
  H.~Li, S.~Suzuki, K.-Y. Wu, and P.~Xie.
\newblock Interleaver technology: Comparisons and applications requirements.
\newblock {\em J. Lightwave Technol.}, 22(1):281, Jan 2004.

\bibitem{Cheng:13}
Chi-Hao Cheng and Shasha Tang.
\newblock Michelson interferometer based interleaver design using classic iir
  filter decomposition.
\newblock {\em Opt. Express}, 21(25):31330--31335, Dec 2013.

\bibitem{1994JLwT...12..471D}
E.~M. {Dowling} and D.~L. {Macfarlane}.
\newblock {Lightwave lattice filters for aptically multiplexed communication
  systems}.
\newblock {\em Journal of Lightwave Technology}, 12:471--486, March 1994.

\bibitem{MacFarlane:94}
Duncan~L. MacFarlane and Eric~M. Dowling.
\newblock Z-domain techniques in the analysis of fabry--perot \'{e}talons and
  multilayer structures.
\newblock {\em J. Opt. Soc. Am. A}, 11(1):236--245, Jan 1994.

\bibitem{Madsen_1998}
C.~K. {Madsen}.
\newblock {Efficient architectures for exactly realizing optical filters with
  optimum bandpass designs}.
\newblock {\em IEEE Photonics Technology Letters}, 10:1136--1138, August 1998.

\bibitem{Ozdemir_2017}
Ahmet~Arda Ozdemir and Suat Gumussoy.
\newblock Transfer function estimation in system identification toolbox via
  vector fitting.
\newblock {\em IFAC-PapersOnLine}, 50(1):6232--6237, 2017.

\bibitem{Pintelon_2012}
Rik Pintelon and Johan Schoukens.
\newblock {\em System identification: a frequency domain approach}.
\newblock John Wiley \& Sons, 2012.

\bibitem{pruessner2013}
Marcel~W Pruessner, Todd~H Stievater, Peter~G Goetz, William~S Rabinovich, and
  Vincent~J Urick.
\newblock Cascaded integrated waveguide linear microcavity filters.
\newblock {\em Applied Physics Letters}, 103(1):011105, 2013.

\bibitem{vandeStadt:85}
Herman van~de Stadt and Johan~M. Muller.
\newblock Multimirror fabry--perot interferometers.
\newblock {\em J. Opt. Soc. Am. A}, 2(8):1363--1370, Aug 1985.

\bibitem{yim2004}
Seongmin Yim and Henry~F Taylor.
\newblock Spectral slicing optical waveguide filters for dense wavelength
  division multiplexing.
\newblock {\em Optics communications}, 233(1-3):113--117, 2004.

\bibitem{yim2005}
Seongmin Yim and Henry~F Taylor.
\newblock Design and spectral characteristics of multireflector etalons.
\newblock {\em Journal of lightwave technology}, 23(3):1419, 2005.

\bibitem{Zhang:17}
Juan Zhang, Dong Hua, Yipeng Ding, and Yang Wang.
\newblock Digital synthesis of optical interleaver based on a solid
  multi-mirror fabry-perot interferometer.
\newblock {\em Appl. Opt.}, 56(36):9976--9983, Dec 2017.

\bibitem{Zhang:10}
Juan Zhang and Xiaowei Yang.
\newblock Universal michelson gires-tournois interferometer optical interleaver
  based on digital signal processing.
\newblock {\em Opt. Express}, 18(5):5075--5088, Mar 2010.

\end{thebibliography}
\end{document}